\documentclass[conference]{IEEEtran}
\usepackage{amsmath,amsfonts}
\usepackage[ruled,vlined]{algorithm2e}
\usepackage{array}
\usepackage[caption=false,font=normalsize,labelfont=sf,textfont=sf]{subfig}
\usepackage{textcomp}
\usepackage{stfloats}
\usepackage{url}
\usepackage{xcolor}
\usepackage{verbatim}
\usepackage{graphicx}
\usepackage{cite}
\usepackage{microtype} % 添加这一行
\usepackage{tabularx, booktabs}
\usepackage[nolist,nohyperlinks]{acronym}
\usepackage[hidelinks]{hyperref}  % <-- hyperref at end
\def\BibTeX{{\rm B\kern-.05em{\sc i\kern-.025em b}\kern-.08em
    T\kern-.1667em\lower.7ex\hbox{E}\kern-.125emX}}
\begin{document}

\title{Optimizing 6G Dense Network Deployment for the Metaverse Using Deep Reinforcement Learning}

\author{Jie Zhang, Swarna Chetty, Qiao Wang, Chenrui Sun, Paul Daniel Mitchell, David Grace, and Hamed Ahmadi  \\
School of Physics, Engineering and Technology, University of York, York YO10 5DD}  
%\author{Paul Daniel Mitchell,~\IEEEmembership{Senior member,~IEEE,}
%\author{Hamed Ahmadi,~\IEEEmembership{Senior member,~IEEE,}
        % <-this % stops a space
% \thanks{J. Zhang, S.B. Chetty, P.D. Mitchell, H. Ahmadi and D. Grace are with School of Physics, Engineering and Technology, University of York, York YO10 5DD, UK (e-mail: jie.zhang@york.ac.uk; swarna.chetty@york.ac.uk; paul.mitchell@york.ac.uk; hamed.ahmadi@york.ac.uk). Q. Wang is with Samsung Research UK, Staines-upon-Thames TW18 4QE, UK (e-mail: qiao1.wang@samsung.com).}% <-this % 

\markboth{}%
{Shell \MakeLowercase{\textit{et al.}}: A Sample Article Using IEEEtran.cls for IEEE Journals}


\maketitle
\begin{acronym}
\acro{IAB}{Integrated Access and Backhaul}
\acro{DRL}{Deep Reinforcement Learning}
\acro{DQN}{Deep Q-Network}
\acro{DQNs}{Deep Q-Networks}
\acro{DDQN}{Double Deep Q-Network}
\acro{SNR}{Signal-to-Noise Ratio}
\acro{MDP}{Markov Decision Process}
\acro{LoS}{Line of Sight}
\acro{QoS}{Quality of Service}
\acro{MINLP}{Mixed Integer Non-Linear Programming}
\acro{mmWave}{Millimeter Wave}
\end{acronym}

\vspace{-0.2in}
\begin{abstract} As the Metaverse envisions deeply immersive and pervasive connectivity in 6G networks, \ac{IAB} emerges as a critical enabler to meet the demanding requirements of massive and immersive communications. \ac{IAB} networks offer a scalable solution for expanding broadband coverage in urban environments. However, optimizing \ac{IAB} node deployment to ensure reliable coverage while minimizing costs remains challenging due to location constraints and the dynamic nature of cities. Existing heuristic methods, such as Greedy Algorithms, have been employed to address these optimization problems. This work presents a novel \ac{DRL} approach for \ac{IAB} network planning, tailored to future 6G scenarios that seek to support ultra-high data rates and dense device connectivity required by immersive Metaverse applications. We utilize \ac{DQN} with action elimination and integrate \ac{DQN}, \ac{DDQN}, and Dueling \ac{DQN} architectures to effectively manage large state and action spaces. Simulations with various initial donor configurations demonstrate the effectiveness of our \ac{DRL} approach, with Dueling \ac{DQN} reducing node count by an average of 12.3\% compared to traditional heuristics. The study underscores how advanced \ac{DRL} techniques can address complex network planning challenges in 6G-enabled Metaverse contexts, providing an efficient and adaptive solution for \ac{IAB} deployment in diverse urban environments. \end{abstract}


\vspace{-0.1in}
\section{Introduction}
\vspace{-0.1in}
\IEEEPARstart{T}{he} rapid convergence of physical and virtual worlds in the Metaverse is reshaping expectations for next-generation connectivity, driving a need for ultra-high data rates, massive device connectivity, and stringent latency requirements \cite{YAQOOB2023100884}. These demands extend well beyond traditional mobile broadband services and encompass diverse immersive applications—ranging from extended reality (XR) to holographic communications—that collectively form the building blocks of a persistent and fully interactive digital realm\cite{10.1016/j.jnca.2024.103828}. To achieve this vision, \ac{IAB} technology has emerged as a cost-effective and flexible solution for expanding coverage in congested urban environments, particularly at \ac{mmWave} and emerging Terahertz frequencies \cite{10552840}. By consolidating access and backhaul functionalities within the same nodes, \ac{IAB} enables multi-hop connections that can better support the robust, low-latency links required for Metaverse applications, paving the way for highly immersive and pervasive digital experiences \cite{10210197}.

However, \ac{IAB} deployments introduce challenges, particularly regarding location constraints, exacerbated in 6G contexts. The use of \ac{mmWave} and potential Terahertz frequencies requires careful planning to mitigate severe path loss and maintain \ac{LoS} conditions. The need for ultra-dense deployments in 6G further complicates node placement within existing urban infrastructure. Effective solutions must optimize \ac{IAB} node placement and performance to ensure efficient network expansion for 6G systems \cite{10210197}.

Although \ac{IAB} deployment with location consideration is an important problem for dense 6G networks, it has not been sufficiently investigated in the literature. The few existing relevant works can be broadly categorized into two main approaches: heuristic algorithms and \ac{DRL}. Heuristic algorithms have been employed to find near-optimal solutions for node deployment:
Raithatha et al. \cite{DBLP:journals/corr/abs-2103-08408} introduced the K-GA model, which integrates K-means clustering with a genetic algorithm to determine gateway locations and maximize backhaul capacity. Similarly, Lai et al. \cite{DBLP:journals/corr/abs-2108-04483} developed an algorithm for joint resource allocation and node placement in multi-hop \ac{IAB} networks, aiming to maximize the sum rate of users while guaranteeing their \ac{QoS} requirements. The authors formulated the task as a \ac{MINLP} and solved it by decomposing it into three subproblems: association optimization using Lagrangian duality, and subchannel and power allocation employing Successive Convex Approximation. However, heuristic methods may struggle with scalability and adaptability in dynamic 6G network environments.
In parallel, \ac{DRL} has emerged as a promising approach to address the dynamic and complex nature of network planning. Li et al. \cite{9794697} proposed a \ac{DRL}-based framework for 3D mobility control of drone base stations, enabling enhanced network connectivity without requiring precise user location data. The authors developed a \ac{DDQN} algorithm to learn the optimal 3D placement of drone base stations based on the spatial distribution of users and the capacity of the backhaul link. Wang et al. \cite{10214526} investigated the deployment of tethered Unmanned Aerial Vehicles (UAVs) in \ac{mmWave} networks using \ac{DRL}. They employed a \ac{DQN} to minimize deployment costs while optimizing the positions of UAVs and terrestrial base stations. Danford et al. \cite{danford2017endtoend} introduced a hierarchical \ac{DRL} architecture for managing Mobile Access Points (MAPs) in dynamic 5G networks. The authors designed a multi-agent \ac{DRL} system where each MAP is controlled by an independent agent, fostering cooperative learning to optimize MAP numbers and positions.
Despite these advances, the interplay between IAB networks and Metaverse-driven services has remained underexplored, particularly regarding the balance between massive connectivity and ultra-high bandwidth demands. Our previous work \cite{info15010019} proposed a two-stage heuristic approach for sustainable \ac{IAB} network planning. However, this approach did not fully consider the impact of initial donor positions on the overall network. The initial placement of donors can significantly influence the efficiency of the entire \ac{IAB} network,  such as affecting the number of required nodes and coverage. 
To bridge this gap, our work makes the following contributions:
\begin{enumerate} 
% \item We propose an advanced \ac{DRL} framework utilizing \ac{DQN}, \ac{DDQN} and Dueling \ac{DQN} architectures, effectively handling large state and action spaces in complex urban \ac{IAB} deployment scenarios crucial for ultra-dense 6G networks.
\item We propose an advanced \ac{DRL} framework utilizing \ac{DQN}, \ac{DDQN} and Dueling \ac{DQN} architectures, with specialized state and action representations in ultra-dense environments. This framework incorporates innovative action elimination strategies and state abstraction techniques to efficiently manage and reduce the complexity of large state and action spaces.
\item We construct a \ac{MDP} model to formulate \ac{IAB} network planning problem, which aims to balance full area coverage and node deployment minimization, enabling optimization of multi-hop \ac{IAB} node placement strategies in dense network environments.
\item We conduct an extensive comparative study, evaluating proposed models Dueling \ac{DQN} and \ac{DDQN} against standard \ac{DQN} and heuristic approaches, demonstrating superior performance in terms of coverage and deployment efficiency across various urban scenarios.
\end{enumerate}
%\vspace{-0.1in}
\section{System Model and Problem formulation}
%\vspace{-0.05in}
\subsection{System Model}
%\vspace{-0.05in}
We consider \ac{IAB} network deployment within an urban environment, aiming to optimize the placement of \ac{IAB} nodes to ensure efficient high-speed wireless coverage. The network operates in the \ac{mmWave} frequency band, known for its high data rate capability but also its challenges with signal propagation.
Before delving into the system model, it is essential to define the key components of the IAB network. An IAB node refers to a wireless node that combines the functionality of an access point and backhaul node. It provides wireless access to end-users while also serving as a wireless backhaul provider to extend the network coverage. On the other hand, a donor node provides wireless backhaul connectivity to the IAB nodes. It acts as a gateway between the IAB network and the core network infrastructure.
\par The system environment is characterized by all upcoming deployed nodes needing to be deployed in potential locations, and these locations are mainly based on existing urban infrastructures such as lamp-posts, traffic lights, and/or bus stops. These locations are crucial as the power infrastructure is readily available, facilitating the power supply for \ac{IAB} nodes. 
To analyze and optimize network deployment, we introduce the communication model \cite{345405} that captures power transmission, antenna gains, path loss, and atmospheric attenuation factors. 
This model is tailored for the \ac{mmWave} band and ensures that the network meets the required \ac{SNR} thresholds for the user data rate requirements. The key equations for the communication model are:
 \vspace{-0.05in}
\begin{align}\small
P_r &= P_{t} + G_{\text{all}} - L_{\text{all}} - N_0, \label{eq:received_power}\  \\
G_{\text{all}} &= G_t + G_r, \label{eq:antenna_gain}\  \\
L_{\text{all}} &= L_r + O_{\text{tr}} + L_{\text{loss}}, \label{eq:total_loss}\  \\
N_0 &= -174 + 10 \log_{10}(W), \label{eq:thermal_noise}
\end{align}
where $P_r$ is the received power, $P_t$ is the transmitted power, $G_{\text{all}}$ is the total antenna gain, $L_{\text{all}}$ is the total signal attenuation, and $N_0$ is the thermal noise power. The total antenna gain $G_{\text{all}}$ is the sum of the transmitter gain $G_t$ and the receiver gain $G_r$. The total signal attenuation $L_{\text{all}}$ accounts for path loss $L_{\text{loss}}$, atmospheric attenuation $O_{\text{tr}}$, and rain-induced attenuation $L_r$. The thermal noise power $N_0$ is calculated based on the bandwidth $W$.
 \vspace{-0.1in}
 \begin{table}[!t]
\caption{Formulation parameters.}
\label{table:multi_hop_formulation_parameters}
\centering
\renewcommand{\arraystretch}{1.2}
\begin{tabularx}{\columnwidth}{>{\hsize=0.5\hsize}X >{\hsize=1.5\hsize}X}
\toprule
\textbf{Symbol} & \textbf{Description} \\
\midrule
\(i\) & Pre-deployment donor location \\ 
\(j\) & Number of potential nodes locations \\ 
\(k\) & Number of grids needing to be covered \\
\(R_o\) & Overhead or required overhead \\
\(I\) & Set of all donor locations \\ 
\(J\) & Set of all potential nodes locations \\ 
\(K\) & Set of the locations on the grid that need to be covered \\
\(u\) & Set of active users in the coverage of a donor/node \\
\(\alpha_j\) & If a candidate location is chosen to deploy node \\
\(C_{ik}\) & Indicates whether grid \(k \in K\) can be covered when a node is deployed at location \(i \in I\) \\
\(C_{jk}\) & Indicates whether grid \(k \in K\) can be covered when a node is deployed at candidate location \(j \in J\) \\
\(Y_{ij}\) & Indicates whether a donor deployed at location\(i\) can provide backhaul when node deploys in candidate location \(j \in J\) \\
\({Y}'_{jn}\) & Indicates whether node deployed at \(j \in J\) can provide backhaul to another node located at \(n \in J\) \\
\({Y}'_{nm}\) & Indicates whether node at location \(n \in J\) can provide backhaul to another node located at \(m \in J\) \\
\(dis_{ij}/dis_{pq}\) & Distance between two candidate nodes or two nodes \(p\) and \(q\), where \(p,q \in I \cup J \cup U\) \\
\(R_{ij}/R_{pq}\) & Data rate between donor \(i\) and node \(j\) or between two nodes \(p\) and \(q\), where \(p,q \in I \cup J \cup U\) \\
\(A_i/A_j/A_x\) & Access data rate when donor/node is deployed in \(i\)/\(j\), or access data rate of a node or a donor \\
\(F_i\) & Fixed data rate of a donor \(i \in I\) \\
\bottomrule
\end{tabularx}
\end{table}
\subsection{Problem Formulation}
%\vspace{-0.05in}
This study aims to optimize the number and positions of node deployments for extensive area coverage through a multi-hop communication strategy. The focus is on minimizing the total number of nodes required to maintain quality of service levels across all considered areas. The backhaul constraint needs to be satisfied so that each deployed node can get sufficient backhaul data rate from the previous deployed nodes or donors.
\subsubsection{Objective}
The primary goal is to minimize the total number of deployed nodes. We introduce a binary decision variable \(\alpha_j \in \{0,1\}\) for each candidate node \(j \in J\), where \(\alpha_j = 1\) indicates that node \(j\) is deployed, and \(\alpha_j = 0\) indicates that it isn't. The objective function is given by:
\vspace{-0.05in}
\begin{equation}
\min \sum_{j \in J} \alpha_j
\end{equation}
 \vspace{-0.1in}
\subsubsection{Access and Backhaul Constraints}
The deployed nodes must cover the whole area, denoted by a grid $k\in K$, by at least one \ac{IAB} donor or node, ensuring the received power at each location meets the minimum $SNR$ threshold, $SNR_0$. Additionally, the network must support efficient data transfer between nodes, considering  data rate from both backhaul and access side. Therefore, coverage constraints are formulated as:
 \vspace{-0.05in}
\begin{equation}\small
\sum_{j \in J} C_{jk} \alpha_j + \sum_{i \in I} C_{ik} \geq 1, \quad \forall k \in K
\end{equation}
\vspace{-0.1in}
\begin{equation}\small
     C_{ik}=\begin{cases}
     1 &  \text{if } SNR(dis_{i,k})>SNR_0 \\
     0 & \text{otherwise} 
    \end{cases}
        \label{eq:donor_coverage}
\end{equation}
 \vspace{-0.1in}
\begin{equation}\small
     C_{jk}=\begin{cases}
     1 &  \text{if } SNR(dis_{j,k})>SNR_0 \\
     0 & \text{otherwise}
    \end{cases}
        \label{eq:node_coverage}
\end{equation}
Constraints below to ensure backhaul connections are only established over feasible links and to deployed nodes:
\begin{align}
 \vspace{-0.05in}
Y_{ij} &\leq \alpha_j, \quad \forall i \in I, \forall j \in J \\
Y'_{jn} &\leq \alpha_n, \quad \forall j, n \in J, n \neq j 
 \vspace{-0.05in}
\end{align}

To ensure effective  multi-hop network communication and avoid countless hops. A common approach to maintain network integrity is that any node's input data rate must exceed its output data rate\cite{bertsekas1992data}. The data rate constraints are formulated as:
%\scriptsize
\begin{align}
F_i -\! R_o (A_i + \sum_{j \in J} Y_{ij} R_{ij} ) &\geq 0, \quad \forall i \in I \\
R_{ij} Y_{ij} - R_o (A_j + \!\!\!\!\sum_{n \in J, n \neq j}\!\!\!\! Y'_{jn} R_{jn}) &\geq 0, \quad \forall i\! \in \!I, \forall j\! \in \!J \\
R_{jn} Y'_{jn} - R_o (A_n + \!\!\!\!\sum_{m \in J, m \neq n}\!\!\!\! Y'_{nm} R_{nm} ) &\geq 0, \quad \forall j, n \in J
\end{align}
\normalsize

All parameters are listed in Table \ref{table:multi_hop_formulation_parameters}.
These constraints ensure that the input data rate to each donor \(i\) is sufficient to meet its own access data rate and the backhaul data rates to nodes it serves; each deployed node \(j\) receives enough data rate from its donors to cover its own access data rate and the data rates to subsequent nodes it serves; and the data rate from node \(j\) to node \(n\) is sufficient to cover node \(n\)'s access data rate and any further nodes served by \(n\). Since \(j\), \(n\) and \(m\) represent all candidate nodes, these constraints ensure that backhaul connections are only established to deployed nodes over feasible links and that the data rate requirements are met throughout the network planning. 
\vspace{-0.1in}  % 可选：调整表格与前后内容的垂直间距

\section{\ac{DRL} and MDP formulation}
We employ \ac{DRL} to address this problem. \ac{DRL} combines the decision-making capabilities of classical reinforcement learning with the representation learning power of deep neural networks. An agent in the \ac{DRL} framework learns to make decisions by observing the state of the environment, taking actions, and receiving feedback in the form of rewards. These rewards guide the agent to discover policies that maximize cumulative future rewards. More specifically, \ac{DRL} is leveraged to find an optimal deployment strategy that minimizes node deployment while ensuring satisfactory \ac{SNR} and adequate backhaul data rates.

We formulate the multi-hop network deployment as an \ac{MDP} in last part, aiming to minimize the number of nodes while maintaining satisfactory access and backhaul data rates. To fully define our \ac{MDP}, we specify the following key components:

\subsubsection{State (S)}
The state captures the deployment status of nodes, data rates, and connectivity within the network:
\begin{itemize}
    \item \textbf{Deployment Status of Nodes} ($D$):
    \[
    D = \begin{bmatrix} d_{11} & d_{12} & \cdots & d_{1n} \\
                        %d_{21} & d_{22} & \cdots & d_{2n} \\
                        \vdots & \vdots & \ddots & \vdots \\
                        d_{n1} & d_{n2} & \cdots & d_{nn} \end{bmatrix} 
    \]
    where $d_{ij}$ is 1 if a node is deployed at location $(i,j)$, and 0 otherwise.

    \item \textbf{Data Rates} ($R$):
    \[
    R = \begin{bmatrix} r_{11} & r_{12} & \cdots & r_{1n} \\
                       %r_{21} & r_{22} & \cdots & r_{2n} \\
                       \vdots & \vdots & \ddots & \vdots \\
                       r_{n1} & r_{n2} & \cdots & r_{nn} \end{bmatrix}
    \]
    where $r_{ij}$ is the data rate left of the node at location $(i,j)$.

    \item \textbf{Node Connectivity Matrix} ($N$):
    \[
    N = \begin{bmatrix}
    n_{11} & n_{12} & \cdots & n_{1n} \\
    %n_{21} & n_{22} & \cdots & n_{2n} \\
    \vdots & \vdots & \ddots & \vdots \\
    n_{n1} & n_{n2} & \cdots & n_{nn}
    \end{bmatrix}
    \]
    where $n_{ij}$ means the number of backhaul connections provided from node $i$ to node $j$, and $n_{ij} = 0$ otherwise.
\end{itemize}

\subsubsection{Actions (A)}
The actions are defined for each potential node location. For each location \(j\), the action can either be 1 for deploying a new node at a potential place or 0 to maintain no change. In this case, the action space size is  \(J\)+1.

\subsubsection{Reward Function (R)}

The reward function is formulated to optimize network deployment by balancing coverage maximization, node deployment minimization, and network connectivity. It is defined as:
%\vspace{-0.1in}
\begin{equation}
\small
R = -\alpha A_{\text{uncov}} - \beta N_{\text{nodes}} + \delta(C)\! -\! \eta \max\left(0, N_{\text{nodes}}\! - \!N_{\text{ref}}\right)
\normalsize
\label{eq:reward_function}
\end{equation}
In \eqref{eq:reward_function}, \( R \) represents the total reward resulting from the agent's action. The term \( -\alpha A_{\text{uncov}} \) penalizes the agent based on the total uncovered area \( A_{\text{uncov}} = \sum_{i \in \mathcal{U}} a_i \), where \( \mathcal{U} \) is the set of uncovered grid indices and \( a_i \) is the area of grid cell \( i \). The coefficient \( \alpha \) determines the weight of this penalty, emphasizing the importance of maximizing coverage.

The second term, \( -\beta N_{\text{nodes}} \), introduces a penalty proportional to the number of deployed nodes \( N_{\text{nodes}} = \sum_{i=1}^{n} d_i \), where \( d_i \) indicates the deployment status of node \( i \) (\( d_i = 1 \) if deployed; \( d_i = 0 \) otherwise). The coefficient \( \beta \) controls the significance of deployment costs, encouraging the agent to minimize the number of nodes used.

The function \( \delta(C) \) provides a coverage reward or penalty based on the achieved coverage percentage \( C \) relative to a predefined threshold \( C_t \):

\begin{equation}
\delta(C) = 
\begin{cases}
-\lambda \left(1 - \dfrac{C}{C_t}\right), & \text{if } C < C_t \\[10pt]
\gamma\, e^{(C - C_t)}, & \text{if } C \geq C_t
\label{eq:reward_function2}
\end{cases}
\end{equation}

In \eqref{eq:reward_function2}, when the coverage \( C \) is below the threshold \( C_t \), a penalty scaled by \( \lambda \) is applied, proportional to the shortfall \( 1 - C/C_t \). When \( C \) meets or exceeds \( C_t \), a reward is granted, increasing exponentially with \( C - C_t \) and scaled by \( \gamma \). This structure motivates the agent to meet or exceed the coverage threshold.

The final term of \eqref{eq:reward_function}, \( -\eta \max\left(0,\; N_{\text{nodes}} - N_{\text{ref}}\right) \), imposes a penalty if the number of deployed nodes exceeds a reference number \( N_{\text{ref}} \), with \( \eta \) controlling the penalty's weight discouraging unnecessary node deployments.

Overall, the reward function guides the agent to maximize coverage while minimizing deployment costs, promoting effective and efficient network deployment strategies. By balancing penalties and rewards through the coefficients \( \alpha \), \( \beta \), \( \lambda \), \( \gamma \), and \( \eta \), the agent is steered toward optimal configurations that meet coverage goals without excessive resource expenditure.
\begin{figure}[ht]\small
\centering
\includegraphics[width=1.05\linewidth]{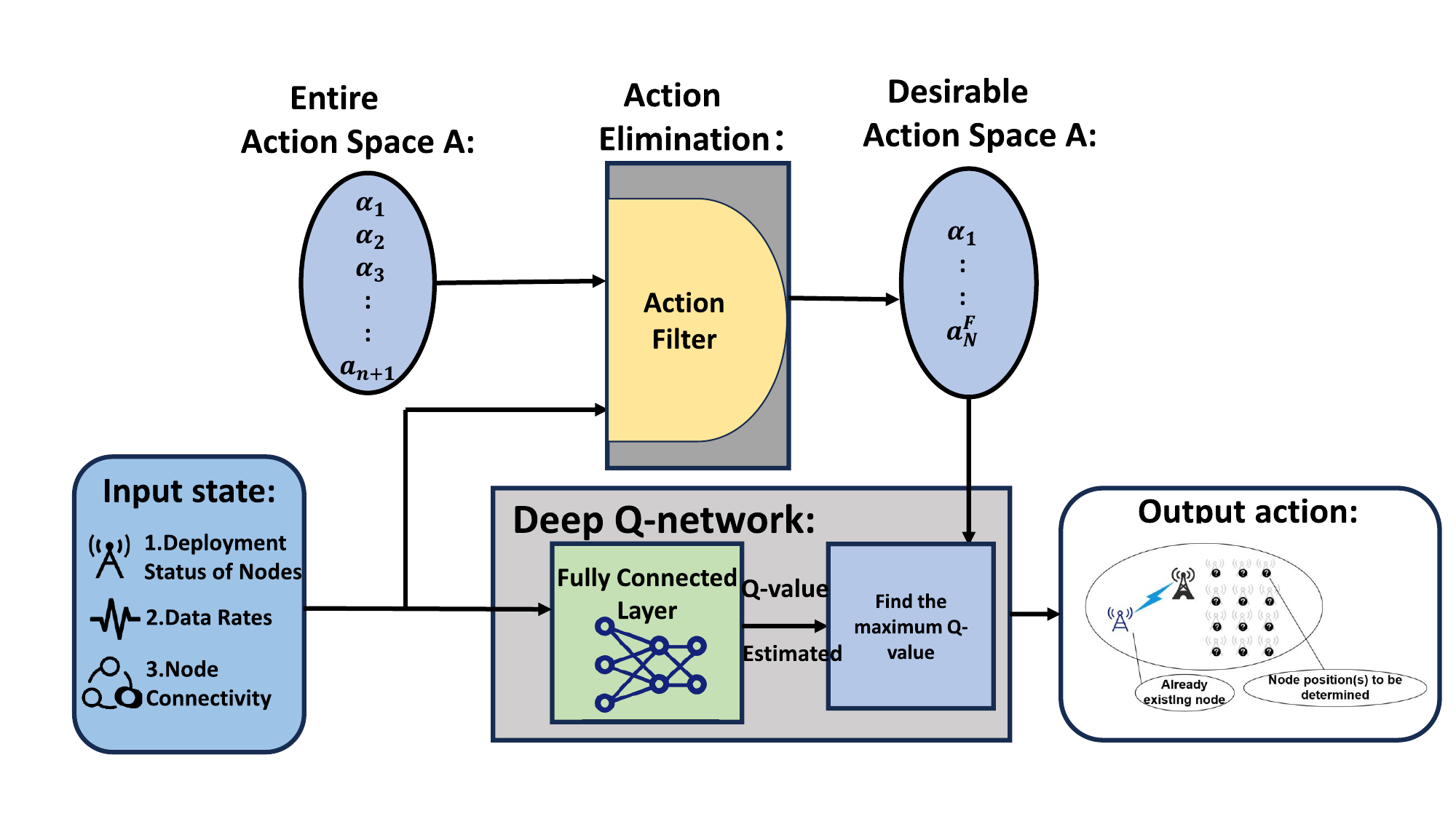}
\caption{Deep Q Network with Action Elimination for IAB Network Planning}
\label{fig:dqn_action_elimination}
\end{figure}

\subsubsection{ \ac{DQNs} and Action Elimination}
 \ac{DQNs} enhance Q-learning by employing neural networks to approximate the optimal action-value function $Q^*(s, a)$ in complex environments. In the \ac{IAB} network planning, \ac{DQNs} are utilized to learn optimal node placement strategies. The overall system architecture, including the integration of \ac{DQN} with action elimination, is illustrated in Figure \ref{fig:dqn_action_elimination}.

The \ac{DQN} approach is characterized by two key equations. The first equation is the \ac{DQN} update rule, which is defined as $Q(s_t, a_t; \theta) = r_t + \gamma \max_{a'} Q(s_{t+1}, a'; \theta^-)$. In this equation, $s_t$ and $s_{t+1}$ represent the current and next states, $a_t$ is the action, $r_t$ is the reward, $\gamma$ is the discount factor, and $\theta$ and $\theta^-$ are the parameters of the Q-network and target network, respectively. The second equation is the training objective, which is expressed as $L(\theta) = \mathbb{E} \left[ \left( y_t - Q(s_t, a_t; \theta) \right)^2 \right]$. Here, $y_t = r_t + \gamma \max_{a'} Q(s_{t+1}, a'; \theta^-)$ is the target Q-value, which is used to calculate the loss function and update the parameters of the Q-network.

 In \ac{IAB} network planning, the action space can be extremely large, as each potential node location is a possible action. However, many actions may be invalid or unnecessary in a given network state, such as deploying nodes in already covered areas. Action elimination prunes these ineffective actions and reduces the search space by dynamically narrowing down the feasible action set based on the current state and constraints. This approach significantly accelerates the convergence of reinforcement learning algorithms and enhances training efficiency. This is implemented through two main algorithms:
\begin{algorithm}[htbp]

\small
\SetAlgoLined
\caption{DQN with Action Elimination for Network Deployment}
\For{episode in episodes}{
    Reset environment and initialize deployed nodes \\
    \While{not done}{
        Obtain valid actions using Algorithm 2 \\
        action = select\_action(state, valid\_actions) \\
        next\_state, reward, done = step(action) \\
        store\_transition(state, action, reward, next\_state, done) \\
        agent.experience\_replay() \\
        state = next\_state \\
    }
    update\_target\_network() \\
    epsilon\_decay() \\
}
\end{algorithm}

\begin{algorithm}[htbp]
\small 
\SetAlgoLined
\caption{Action filter}
\KwIn{All actions, Deployed nodes, taken actions, min distance}
\KwOut{filtered\_actions}
Initialize filtered actions \\
\For{action in All actions}{
    \If{action is in deployed nodes}{
         \quad 
    }
check whether action meet backhaul constraint \\
    \For{left actions}{
     calculate distance between new action and taken actions) \\
        \If{distance $>$ min distance}{
            Action is not valid 
        }
    }
    \If{is valid}{
        Add action to filtered\_actions \\
    }
}
\Return filtered actions
\end{algorithm}
% \vspace{-0.15in}
Algorithm 1 provides a high-level overview of our DQN with action elimination approach, while Algorithm 2 details the action filtering process that is crucial for reducing the action space.
This approach enables the \ac{DQN} to learn efficient network deployment strategies by focusing on feasible and promising actions, leading to improved coverage with fewer nodes in our problem.

 \vspace{-0.1in}
\section{Simulation Result and Analysis}

To evaluate the performance and adaptability of different reinforcement learning models, we first implement the algorithms described in Section II on \ac{DQN}, And then expanding to \ac{DDQN}, and Dueling \ac{DQN} and compare them with a heuristic approach. To verify the robustness of our algorithms in different initial donor environments, we test three distinct initial donor placement patterns: a five-dice pattern for balanced distribution, a vertical pattern for linear arrangement, and a pentagon pattern for dispersed geometric distribution. By simulating these diverse scenarios, we aim to gain comprehensive insights into how these models perform and adapt to various initial network configurations. This approach assesses the flexibility and effectiveness of our proposed methods in real-world urban environments with different donor placement constraints. Table \ref{tab:dqn_key_parameters} summarizes the key simulation parameters, network settings, and deep network settings used across all three initial donor configurations.
\begin{table}[ht]
    \centering
    \caption{DQN Architecture and Key Hyperparameters}
    \label{tab:dqn_key_parameters}
    \begin{tabular}{@{}ll@{}}
        \toprule
        \textbf{Parameter}                     & \textbf{Value}                                \\ \midrule
        \textbf{Neural Network Architecture}   &                                               \\ \midrule
        Hidden Layers                          & 3 Layers                                      \\
        Neurons per Layer                      & 1024, 512, 256                                \\
        Activation Function                    & ReLU                                          \\
        Output Layer                           & Outputs $n_{\text{actions}}$ Q-values          \\
        Layer Normalization                    & After each hidden layer                       \\ \midrule
        \textbf{Training Hyperparameters}      &                                               \\ \midrule
        Learning Rate ($\alpha$)               & 0.001                                         \\
        Optimizer                              & Adam                                          \\
        Discount Factor ($\gamma$)             & 0.99                                                              \\
        Batch Size                             & 512                                           \\
        Replay Memory Size                     & 20,000 transitions                            \\
        Target Network Update Frequency        & Every 64 steps                                \\ \midrule
        \textbf{Environment Parameters}        &                                               \\ \midrule
        Map Size                               & 1000m $\times$ 1000m                           \\
        Grid Size                              & 50m                                           \\
        Node Coverage Radius                   & 200m                                          \\
        Node Backhaul Radius                   & 300m                                          \\
        Frequency Band                         & 60 GHz                                        \\
        Overhead Factor                        & 1.2                                           \\
        Node Data Rate                         & 2 Gbps                                        \\
        Donor Data Rate                        & 30 Gbps                                       \\ \bottomrule
    \end{tabular}
\end{table}
 \vspace{-0.2in}
\subsection{Models Compared}

In this study, we compare several algorithms, starting with a heuristic algorithm from our previous work \cite{info15010019}, which uses a greedy strategy to maximize coverage while ensuring network connectivity. The main model proposed is based on \ac{DQN}. To address overestimation issues in \ac{DQN}, we also explore \ac{DDQN}, which uses separate networks for action selection and evaluation, leading to more stable learning. Additionally, the Dueling DQN model is introduced, which improves performance by separating state value and action advantage estimation.
\begin{enumerate}
\item Heuristic Algorithm: A greedy strategy from our previous work \cite{info15010019} serves as a baseline that iteratively selects locations providing maximum coverage increase while ensuring network connectivity. 
\item \ac{DQN}: The main model proposed in this study. Subsequent models are based on \ac{DQN}.
\item \ac{DDQN}: Addresses overestimation problem in \ac{DQN} by using separate networks for action selection and evaluation, leading to more stable learning.
\item Dueling \ac{DQN}: Separates state value and action advantage estimation in \ac{DQN}, allowing better generalization and improved performance in states where action choice has less impact.
\end{enumerate}
 \vspace{-0.1in}
\subsection{Simulation Result}
\begin{figure}[!htbp]
    \centering
    \includegraphics[width=0.95\columnwidth,height=0.2\textheight]{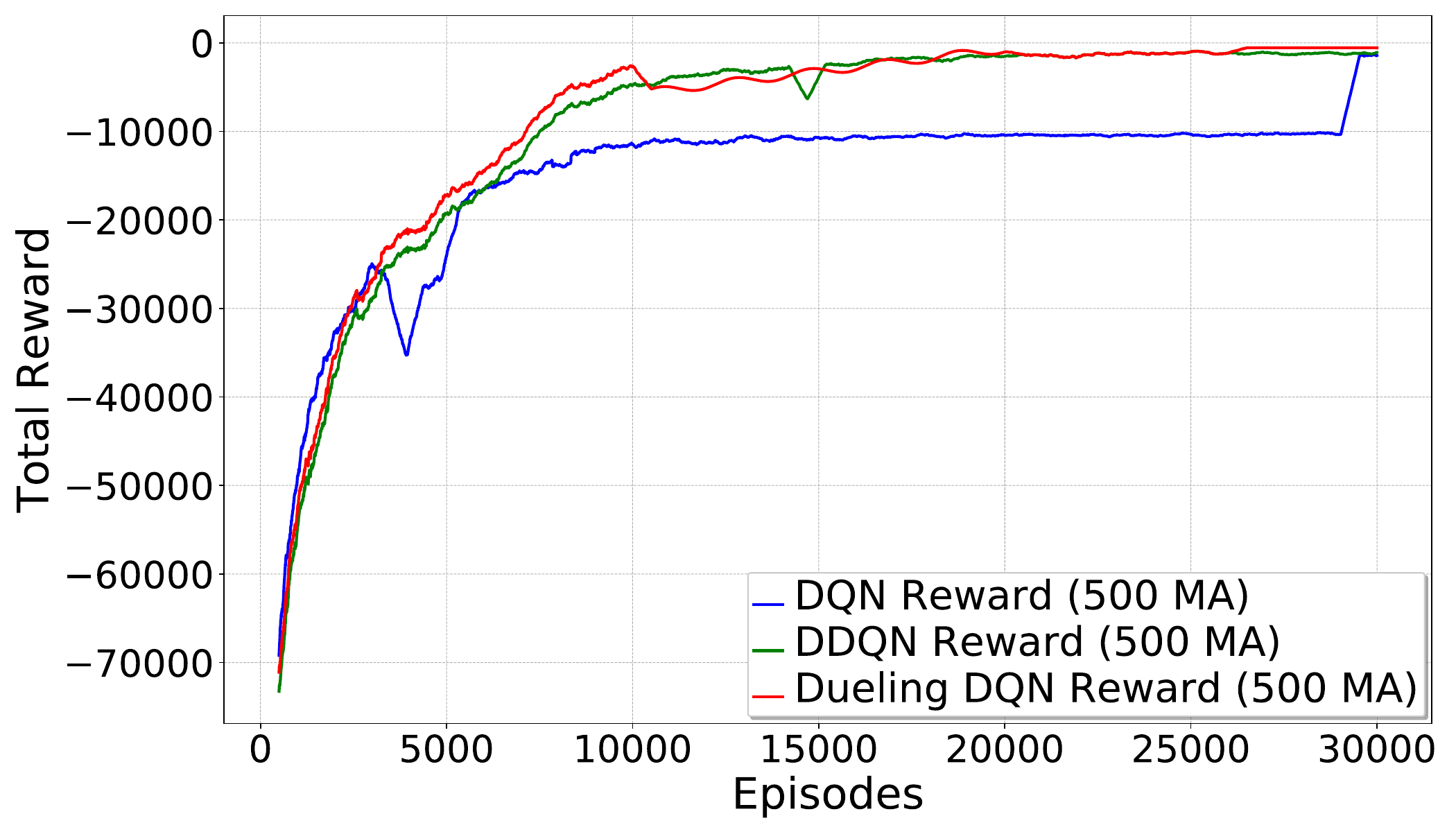}
     \vspace{-0.1in}
    \caption{ Reward vs Episodes Comparison in five-dice distribution donor environment.}
    \label{fig:reward_vs_episodes}
\end{figure}

\begin{figure}[!htbp]
    \centering
	\includegraphics[width=0.7\columnwidth]{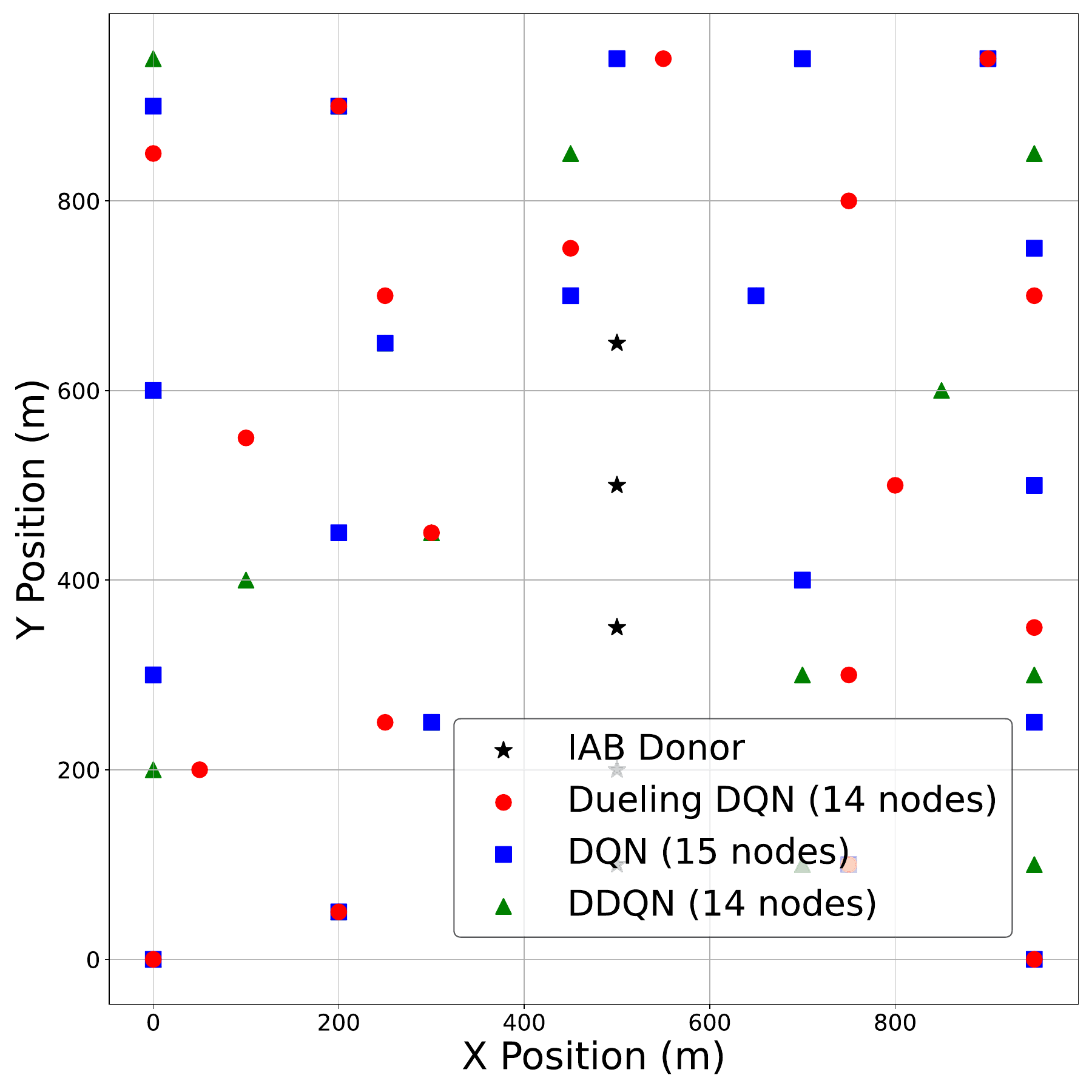}
    \caption{ Final network planning for three models in vertical distribution donor environment.}
    \label{fig:network_planning}
\end{figure}

To investigate the learning behavior of the reinforcement learning models, we examine the convergence of the reward over the training episodes. Figure \ref{fig:reward_vs_episodes} presents the reward convergence plot for \ac{DQN}, \ac{DDQN}, and Dueling \ac{DQN} with 500 moving windows in five-dice environment. All three models demonstrate a consistent improvement in the reward as the training progresses, indicating their ability to learn effective deployment strategies. However, the rate of convergence and the final reward values differ among the models. Dueling \ac{DQN} demonstrates the fastest convergence, followed by \ac{DDQN}, and then \ac{DQN}. This order of convergence speed can be attributed to the architectural differences among the models. Notably, while \ac{DQN} initially shows slower convergence, it eventually achieves the same reward level as \ac{DDQN} and Dueling \ac{DQN} after approximately 29000 episodes.
     \vspace{-0.1in}
To test the performance of the proposed models in different initial donor environments, we conduct experiments with donors arranged in a vertical pattern. Figure \ref{fig:network_planning} illustrates the best deployment test results from 100 tests for \ac{DQN}, \ac{DDQN}, and Dueling \ac{DQN} in this setting. And both Dueling \ac{DQN} and \ac{DDQN} achieve full coverage with least nodes cost. 

\begin{figure}[!htbp]
    \centering
    \includegraphics[width=\columnwidth,height=0.2\textheight]{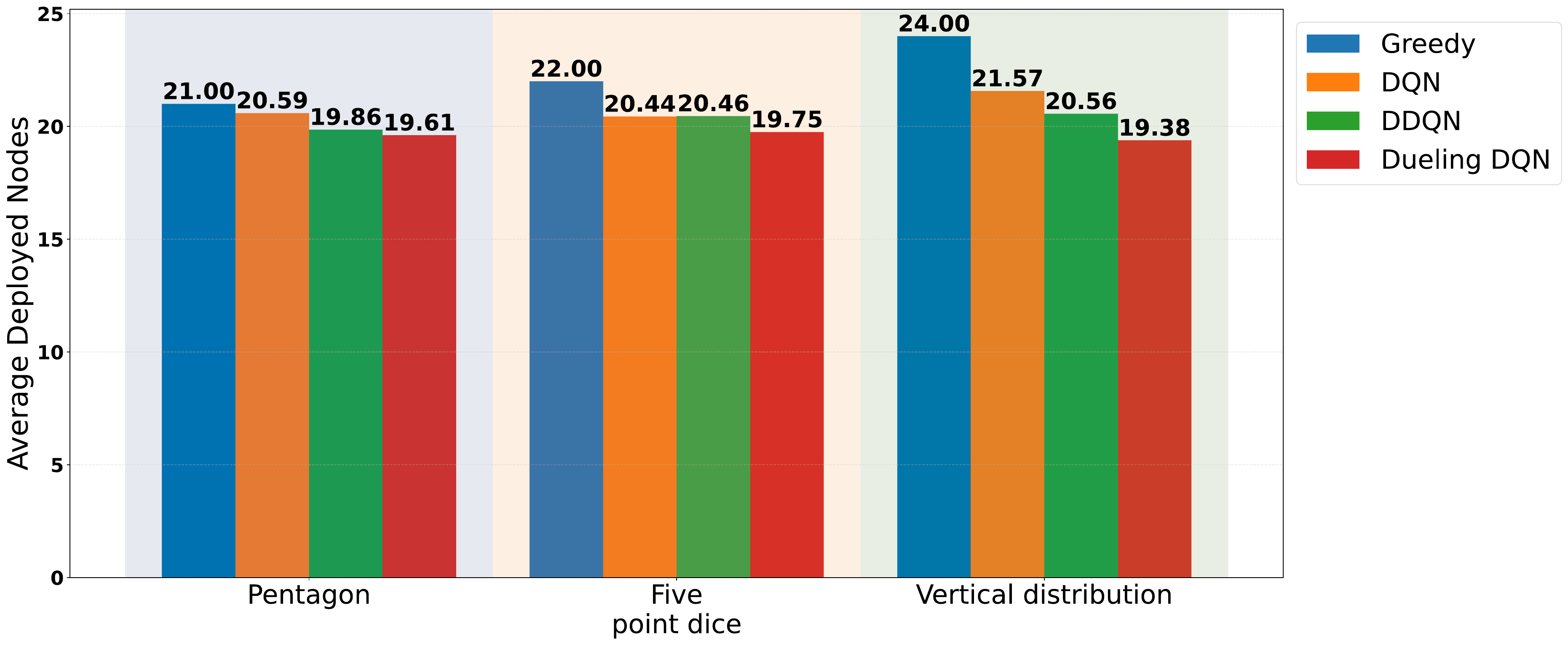}
    \caption{Deployed Nodes vs Different Initial Donor Environment.}
    \label{fig:deployed_nodes}
\end{figure}
Figure \ref{fig:deployed_nodes} compares the average number of nodes deployed by a heuristic approach (Greedy) and three reinforcement learning models (\ac{DQN},\ac{DDQN}, and Dueling \ac{DQN}) across three environments. The Greedy approach consistently requires the most nodes (mean 22.33) and shows the highest variance, indicating poor adaptability. In contrast, the \ac{DRL} models demonstrate superior performance and flexibility. Dueling \ac{DQN} emerges as the most efficient, deploying the fewest nodes on average (19.58) and maintaining the most consistent performance across environments, suggesting excellent generalization capabilities. \ac{DDQN} (mean 20.29) and \ac{DQN} (mean 20.87) also show significant improvements over Greedy, with \ac{DDQN} generally outperforming \ac{DQN}. These results suggest that Dueling \ac{DQN}'s architecture enables it to learn more efficient and generalizable deployment strategies, offering a promising solution for adaptive \ac{IAB} network deployment in diverse urban settings. Future work will focus on incorporating transfer learning techniques to leverage knowledge across different urban environments, as well as integrating dynamic user mobility patterns to further enhance the adaptability and efficiency of our \ac{DRL}-based \ac{IAB} network deployment framework.

\vspace{-0.15in}
\section{Conclusion}
\vspace{-0.05in}
This study presents an innovative approach to optimizing \ac{IAB} network deployment in urban environments using DRL. By formulating the problem as a MDP and employing \ac{DQN} with action elimination, our method learns efficient node placement strategies that minimize deployed nodes while ensuring full coverage and sufficient backhaul quality service. The proposed algorithms, particularly the \ac{DQN} with action elimination (Algorithms 1 and 2), have shown promising results in addressing the complex challenges of IAB network deployment for future 6G systems.
The research compares the performance of \ac{DQN}, \ac{DDQN}, and Dueling \ac{DQN} against a heuristic greedy approach across different initial donor environments. Simulation results demonstrate the superiority of the \ac{DRL} approach, with Dueling \ac{DQN} exhibiting the fastest convergence and most efficient deployment. This framework offers a promising solution for efficient and adaptive \ac{IAB} network deployment in urban settings, potentially reducing infrastructure costs while maintaining high quality of service. 
\par The study's findings underscore the potential of advanced machine learning techniques in addressing complex network planning challenges, paving the way for more intelligent and cost-effective 6G and beyond network deployments that are critical for the realization of the Metaverse. In future work, we plan to extend our \ac{DRL}-based approach to the Open Radio Access Network (ORAN) architecture. ORAN’s disaggregated, vendor-neutral design and real-time monitoring capabilities can be leveraged to incorporate additional network metrics—such as traffic load and interference—into our state representations. By integrating our algorithms into ORAN, we aim to enhance network flexibility and interoperability, further supporting the dynamic and scalable network requirements of real-world 6G networks and the evolving Metaverse ecosystem.
\bibliographystyle{IEEEtran}

\end{document}